\documentclass[a4paper]{jpconf}
\usepackage{graphicx}
\begin{document}
\title{Magnetic fields and gas flows around circumnuclear starbursts}

\author{Rainer Beck}

\address{Max-Planck-Institut f\"ur Radioastronomie, Auf dem
H\"ugel 69, 53121 Bonn, Germany}

\ead{rbeck@mpifr-bonn.mpg.de}

\begin{abstract}
Radio continuum observations of barred galaxies revealed strong
magnetic fields of $\ge50-100\,\mu$G in the circumnuclear
starbursts. Such fields are dynamically important and give rise to
magnetic stress that causes inflow of gas towards the center at a
rate of several solar masses per year, possibly along the spiral
field seen in radio polarization and as optical dust lanes. This may
solve the long-standing question of how to feed active nuclei, and
explain the relation between the bolometric luminosity of AGN nuclei
and the star-formation rate of their hosts. The strong magnetic
fields generated in young galaxies may serve as the link between
star formation and accretion onto supermassive black holes. --
Magnetic fields of $\ge160\,\mu$G strength were measured in the
central region of the almost edge-on starburst galaxy NGC~253. Four
filaments emerging from the inner disk delineate the boundaries of
the central outflow cone of hot gas. Strong Faraday rotation of the
polarized emission from the background disk indicates a large-scale
helical field in the outflow walls.
\end{abstract}

\section{Introduction}

Magnetic fields are a major agent in the ISM and also control the
density and propagation of cosmic rays. The radio--infrared
correlation indicates that turbulent fields are strongest in
star-forming regions. Magnetic fields and cosmic rays can provide
the pressure to drive galactic outflows. Outflows from starburst
galaxies in the early Universe may have magnetized the intergalactic
medium. In spite of our increasing knowledge of cosmic magnetic
fields, many important questions, especially their origin, strength
in intergalactic space, first occurrence in young galaxies and their
dynamical importance for galaxy evolution remain unanswered.

\section{Origin of galactic magnetism}

The most promising mechanism to sustain magnetic fields in the
interstellar medium of galaxies is the {\em dynamo}. In young
galaxies a small-scale dynamo probably amplified seed fields from
the protogalactic phase to the energy density level of turbulence
within less than $10^9$~yr (Schleicher et al. \cite{schleicher10}).
To explain the generation of large-scale fields in galaxies, the
mean-field dynamo has been developed (Beck et al. \cite{beck96}). It
is based on turbulence, differential rotation and helical gas flows,
driven by supernova explosions (Gressel et al. \cite{gressel08}) and
cosmic rays (Hanasz et al. \cite{hanasz09}). The mean-field dynamo
in galaxy disks predicts that within a few $10^9$~yr large-scale
regular fields are excited from the seed fields (Arshakian et al.
\cite{arshakian09}), forming patterns (``modes'') with different
azimuthal symmetries in the disk and vertical symmetries in the
halo.

\section{Observation of galactic magnetism}

Most of what we know about galactic magnetic fields comes through
the detection of radio waves. The intensity of {\em synchrotron
emission}\ is a measure of the number density of cosmic-ray
electrons and of the strength of the total magnetic field component
in the sky plane. The assumption of energy equipartition between the
total cosmic rays and total magnetic fields allows us to calculate
the total magnetic field strength from the synchrotron intensity
(Beck \& Krause \cite{beck+krause05}).

Polarized emission originates from ordered fields. As polarization
``vectors'' are ambiguous by $180^\circ$, they cannot distinguish
{\em regular (coherent) fields}, defined to have a constant
direction within the telescope beam, from {\em anisotropic fields},
which are generated from turbulent fields by compressing or shearing
gas flows, so that their direction frequently reverses perpendicular
to the flow direction. Unpolarized synchrotron emission indicates
{\em turbulent fields}\, which have random directions.

The intrinsic degree of linear polarization of synchrotron emission
is about 75\%. The observed degree of polarization is smaller due to
the contribution of unpolarized thermal emission, by {\em Faraday
depolarization}\ along the line of sight and across the beam
(Sokoloff et al. \cite{sokoloff98}) and by geometrical
depolarization due to variations of the field orientation within the
beam.

At radio wavelengths of a few centimeters and below, the orientation
of the observed B--vector is parallel to the field orientation, so
that the magnetic patterns of many galaxies can be mapped directly
(e.g. Beck \cite{beck05}). At longer wavelengths, the observed
polarization vector is rotated in a magnetized thermal plasma by
{\em Faraday rotation}. The rotation angle increases with the square
of the wavelength $\lambda^2$ and with the {\em Rotation Measure
(RM)}, which is the integral of the plasma density and the strength
of the component of the field along the line of sight. As the
rotation angle is sensitive to the sign of the field direction, only
regular fields give rise to Faraday rotation, while anisotropic and
random fields do not.

\section{Magnetic fields in barred galaxies}

\begin{figure*}[t]
\vspace*{2mm}
\begin{minipage}[t]{8cm}
\begin{center}
\includegraphics[bb = 48 49 522 552,width=8cm,clip=]{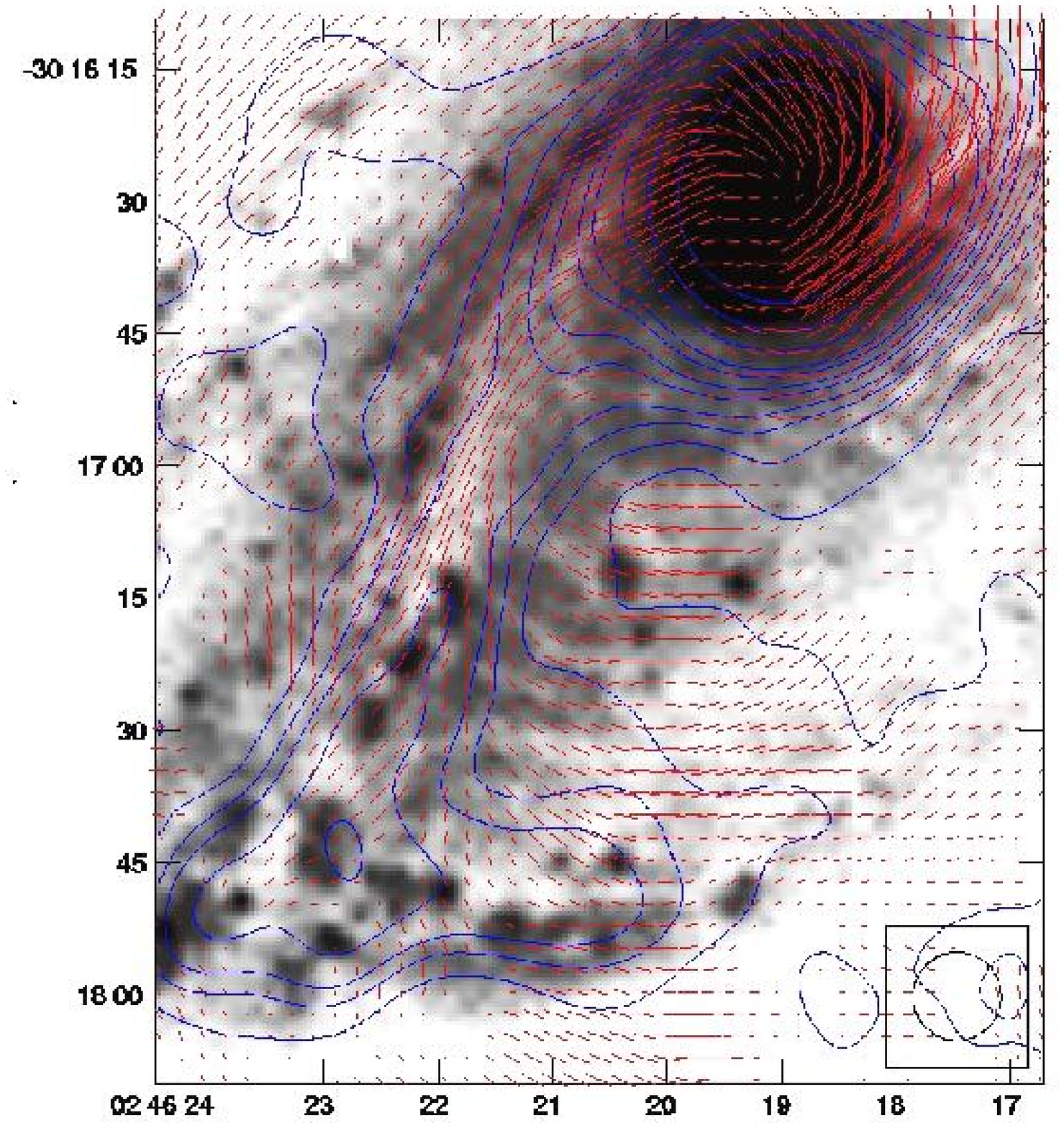}
\caption{Total radio emission (contours) and $B$--vectors of the
barred galaxy NGC~1097, observed at 6~cm wavelength with the {\it
VLA}\ and smoothed to 10'' resolution (Beck et al. \cite{beck05a}).
The background optical image is from Halton Arp ({\it Cerro Tololo
Observatory}).} \label{fig:n1097}
\end{center}
\end{minipage}\hfill
\begin{minipage}[t]{7cm}
\begin{center}
\includegraphics[bb = 50 51 519 503,width=7cm,clip=]{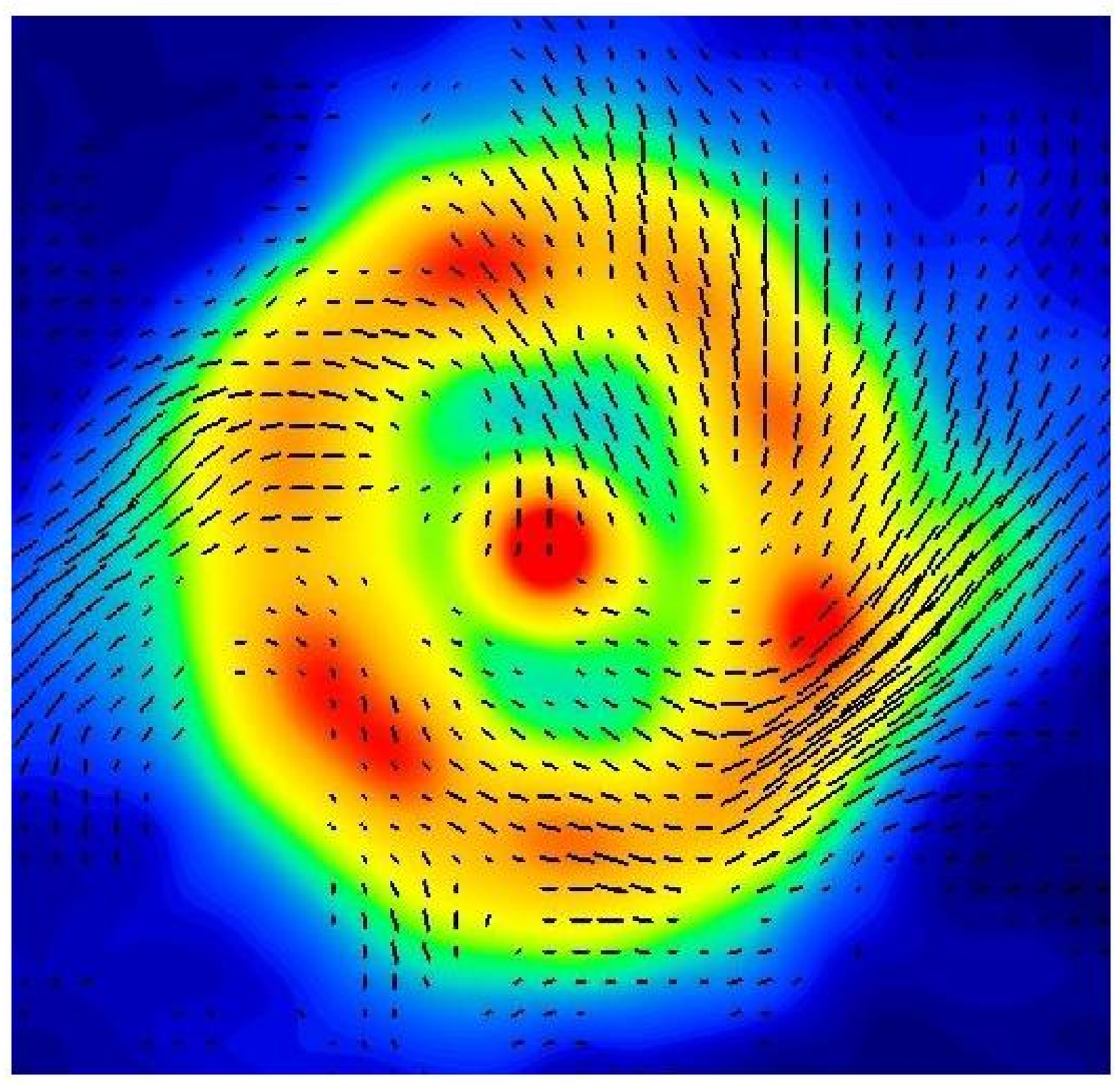}
\caption{Total radio emission (colors) and $B$--vectors of the
circumnuclear ring of the barred galaxy NGC~1097, observed at 3~cm
wavelength with the {\it VLA}\ and smoothed to 3'' resolution (Beck
et al. \cite{beck05a}).} \label{fig:n1097_center}
\end{center}
\end{minipage}
\end{figure*}

The gravitational potential of strongly barred galaxies drives
noncircular orbits and gas inflow. In many barred galaxies a
circumnuclear ring is formed. Numerical models show that gas
streamlines are deflected in the bar region along shock fronts,
behind which the cold gas is compressed in a fast shearing flow
(Athanassoula \cite{athanassoula}). The compression regions traced
by massive dust lanes develop along the edge of the bar that is
leading with respect to the galaxy's rotation because the gas
rotates faster than the bar pattern.

20 galaxies with large bars were observed with the Very Large Array
(VLA) and with the Australia Telescope Compact Array (ATCA) (Beck et
al. \cite{beck02,beck05a}). The total radio luminosity (a measure of
the total magnetic field strength) is strongest in galaxies with
high far-infrared luminosity (indicating high star-formation
activity), a result similar to that in non-barred galaxies. The
average radio intensity, radio luminosity and star-formation
activity all correlate with the relative bar length. Polarized
emission was detected in 17 of the 20 barred galaxies.

NGC~1097 (Fig.~\ref{fig:n1097}) is one of the nearest barred
galaxies and hosts a huge bar of about 16\,kpc length. The total and
polarized radio intensities are strongest in the downstream region
of the dust lanes (southeast of the center) by compression of
turbulent fields in the bar's shock.
The pattern of field lines in NGC~1097 is similar to that of the gas
streamlines as obtained in numerical simulations (Athanassoula
\cite{athanassoula}). This suggests that the ordered magnetic field
is aligned with the flow and amplified by shear. Remarkably, the
optical image of NGC~1097 shows dust filaments in the upstream
region which are aligned with the ordered field
(Fig.~\ref{fig:n1097}). Between the region upstream of the southern
bar and the downstream region the field lines smoothly change their
orientation by almost $90^\circ$. The ordered field is also hardly
compressed, probably coupled to the diffuse gas and strong enough to
affect its flow (Beck et al. \cite{beck05a}). The polarization
pattern in barred galaxies can be used as a tracer of shearing gas
flows in the sky plane and complements spectroscopic measurements of
radial velocities.

\section{Magnetic fields in circumnuclear rings and gas inflows}

Barred galaxies can create a gas reservoir and starburst in the
central kiloparsec, fuelling an AGN needs other processes (e.g.
Davies et al. \cite{davies10}). The central regions of barred
galaxies are often sites of ongoing intense star formation and
strong magnetic fields. Bright radio emission has been observed in
several galaxies with the following equipartition strengths of the
total magnetic fields (Beck et al. \cite{beck05b}):

\begin{itemize}
\item NGC~1097: $\simeq50$\,$\mu$G in the circumnuclear ring,
50\,$\mu$G in the ring knots
\item NGC~1365: $\simeq60$\,$\mu$G in the dust lanes (no ring)
\item NGC~1672: $\simeq70$\,$\mu$G in the ring knots
\item NGC~7552: $\simeq105$\,$\mu$G in the ring
\end{itemize}

Note that these values are lower limits because the energy densities
of cosmic-ray protons and electrons in starburst regions are reduced
by various losses (Thompson et al. \cite{thompson06}). This is
supported by an investigation of the $\gamma$-ray and radio emission
from the starburst galaxies M~82 and NGC~253 which suggests that
most of the radio emission is from secondaries undergoing strong
bremsstrahlung and ionization losses (Lacki et al. \cite{lacki11}).

One of the key problems of AGN physics is how to fuel the nucleus
with gas from the central region, which needs outwards transport of
angular momentum (Davies et al. \cite{davies09}). Magnetic fields
can help here. The circumnuclear ring of NGC~1097 is bright in the
optical, IR and radio spectral ranges. It has a diameter of about
1.5\,kpc and an active nucleus in its center
(Fig.~\ref{fig:n1097_center}). The nonthermal and weakly polarized
radio emission is a signature of strong turbulent magnetic fields
(Beck et al. \cite{beck05a}). Magnetic stress in the differentially
rotating circumnuclear ring can drive mass inflow (Balbus \& Hawley
\cite{balbus98}) at a rate of:

\begin{eqnarray}
dM/dt = 2\pi \, \sigma \, T_{r\phi} \, / \, \Omega
\end{eqnarray}
where $\sigma$ is the surface mass density, $\Omega$ its angular
rotation velocity and $T_{r \, \phi}$ is the magnetic stress tensor.
$r$ and $\phi$ denote the radial and azimuthal field components. Its
dominant component can be written in terms of the Alfv{\'e}n
velocity $v_A$ as $T_{r\phi} \simeq - < v_{A,r} \, v_{A,\phi} >$.
This yields:

\begin{eqnarray}
dM/dt = - (<b_r \, b_{\phi}> + B_r \, B_{\phi}) \, h \, / \, \Omega
\end{eqnarray}
where $h$ is the scale height of the gas, $b$ the strength of the
turbulent field and $B$ that of the ordered field. The correlation
between $b_r$ and $b_{\phi}$ is generated by shear from differential
rotation. For NGC~1097, $h\simeq100$\,pc, $v\simeq450$\,km/s at
1\,kpc radius, $b_r \simeq b_\phi \simeq$\,50\,$\mu$G and $B_r
\simeq B_\phi \simeq$\,10\,$\mu$G give an inflow rate of several
$M_\odot$/yr, which is sufficient to fuel the activity of the
nucleus of this galaxy (Beck et al. \cite{beck05a}). This mechanism
can be expected to operate in many galaxies.

Equipartition between the energy densities of cosmic rays and
magnetic fields yields:

\begin{eqnarray}
dM/dt \simeq - B_{tot}^2 \, \, h / \, \Omega \, \, \propto -
I_{sync}^{0.5} \, h \, / \, \Omega \, \, \propto - I_{IR}^{0.5} \, h
\, / \, \Omega \, \, \propto - SFR^{0.5} \, h \, / \, \Omega
\end{eqnarray}
where $B_{tot}$ is the total field strength, $I_{sync}$ is the radio
synchrotron intensity which is proportional to the infrared
intensity $I_{IR}$ (the radio-infrared correlation) and hence to the
star-formation rate $SFR$. The relation between the bolometric
luminosity of AGNs $L_{bol}$ and the $SFR$ of their hosts (e.g.
Trakhtenbrot \& Netzer \cite{trakhtenbrot10}) can be explained by
this scenario. The relation between the accretion rate and SFR
(Eq.~3) should be tested with a sample of galaxies.

The ordered field in the ring of NGC~1097 has a spiral pattern and
extends towards the nucleus. Its pitch angle agrees with that of the
spiral filaments seen in dust and H$_2$ (Prieto et al.
\cite{prieto05}, Davies et al. \cite{davies09}), suggesting gas
inflow along the magnetic field lines.

Radio emission is a tracer of star formation also in distant
galaxies. The radio--infrared correlation holds to redshifts of at
least $z\simeq3$ (Murphy \cite{murphy09}), indicating strong
magnetic fields in young galaxies, probably generated by the
turbulent dynamo mechanism (Schleicher et al. \cite{schleicher10}).
Infrared observations indicate that star formation in the hosts of
the most luminous AGNs peaked at $z\simeq3$ and decreased at later
epochs (Serjeant et al. \cite{serjeant10}), which is consistent with
the evolution of the accretion rate onto supermassive black holes
(Trakhtenbrot et al. \cite{trakhtenbrot11}). The strong magnetic
fields in young galaxies may serve as the link between star
formation and accretion.

\section{The nuclear outflow of the starburst galaxy NGC~253}

\begin{figure*}[t]
\vspace*{2mm}
\begin{minipage}[t]{7cm}
\begin{center}
\includegraphics[bb = 19 19 593 570,width=7cm,clip=]{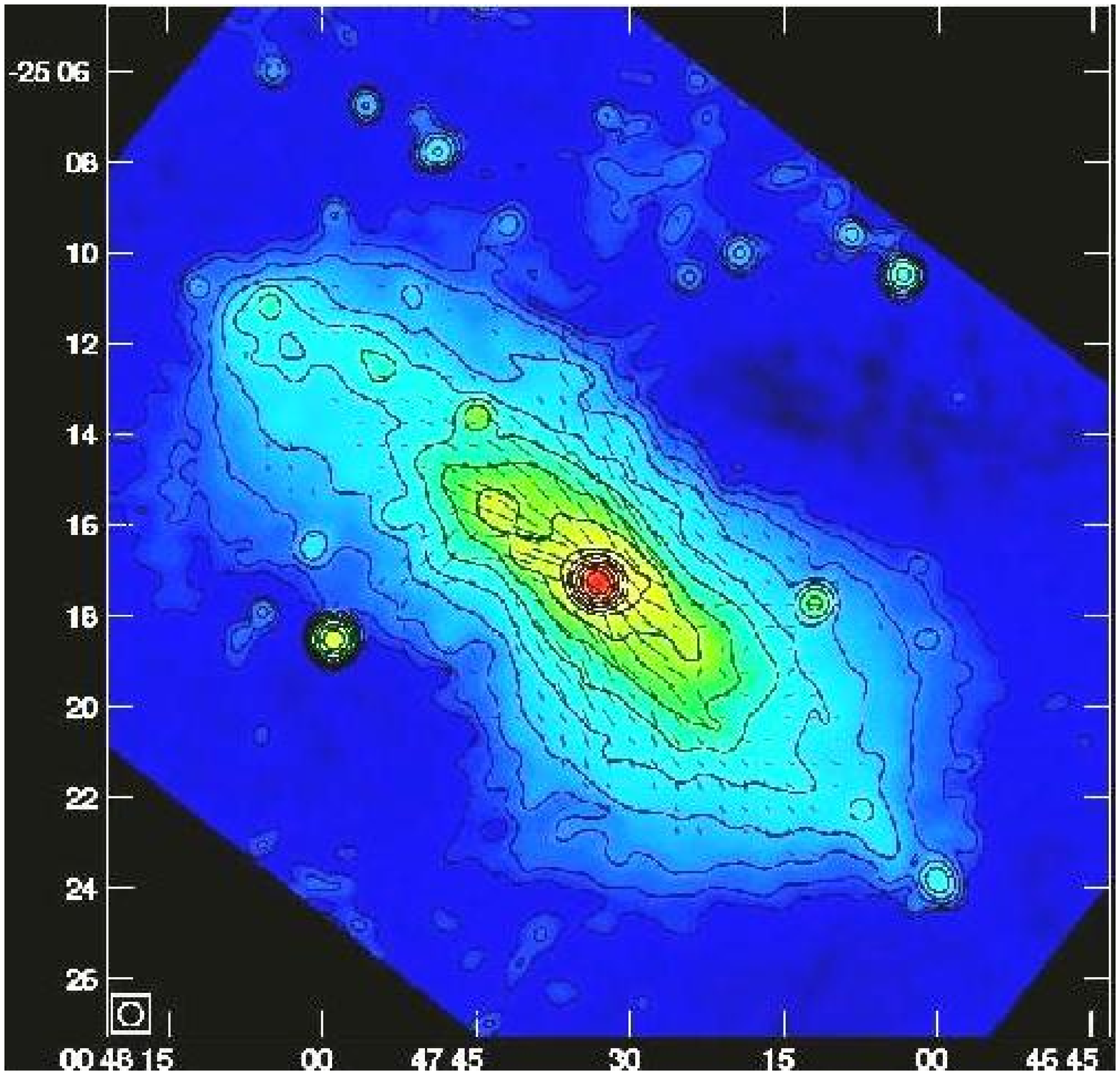}
\caption{Total radio emission (contours and colors) and polarization
$B$--vectors of the spiral galaxy NGC~253, combined from
observations at 6~cm wavelength with the {\it VLA}\ and the {\it
Effelsberg}\ telescope and smoothed to 30'' resolution (Heesen et
al. \cite{heesen09}).} \label{fig:n253}
\end{center}
\end{minipage}\hfill
\begin{minipage}[t]{8cm}
\begin{center}
\includegraphics[bb = 30 30 198 173,width=8cm,clip=]{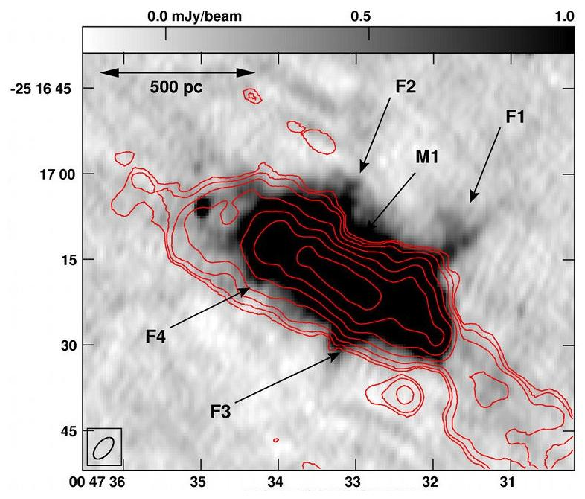}
\caption{Total radio emission (greyscale) of the inner starburst
region of NGC~253, observed at 20~cm with the {\it VLA}\ at a
resolution of 1.3'' x 2.2'' (Heesen et al. \cite{heesen11}). The
radio filaments (F1--4) and the minimum (M1) are indicated. The
contours show the 12CO(2--1) emission, observed with the {\it SMA}
(Sakamoto et al. \cite{sakamoto06}).} \label{fig:n253_center}
\end{center}
\end{minipage}
\end{figure*}

NGC~253 is a prototypical starburst galaxy, hosting a starburst
nucleus (Brunthaler et al. \cite{brunthaler09}). Its high
inclination of $78.5^{\circ}$ allows observing the extraplanar
emission. The bright radio halo (Fig.~\ref{fig:n253}) is probably a
result of a galactic wind with a speed of about 300\,km/s (Heesen et
al. \cite{heesen09}).

Radio observations of the inner starburst region with high
resolution revealed four filaments with widths of less than 40\,pc
and lengths of at least 500\,pc (Fig.~\ref{fig:n253_center}). These
are interpreted as the boundaries of the outflow cone of hot gas
interacting with the halo gas (Heesen et al. \cite{heesen11}). The
equipartition strength of the total field in the filaments is at
least 40\,$\mu$G, that in the central region $\ge160\,\mu$G.

\begin{figure*}[t]
\vspace*{2mm}
\begin{minipage}[t]{8.5cm}
\begin{center}
\includegraphics[bb = 14 14 566 428,width=8cm,clip=]{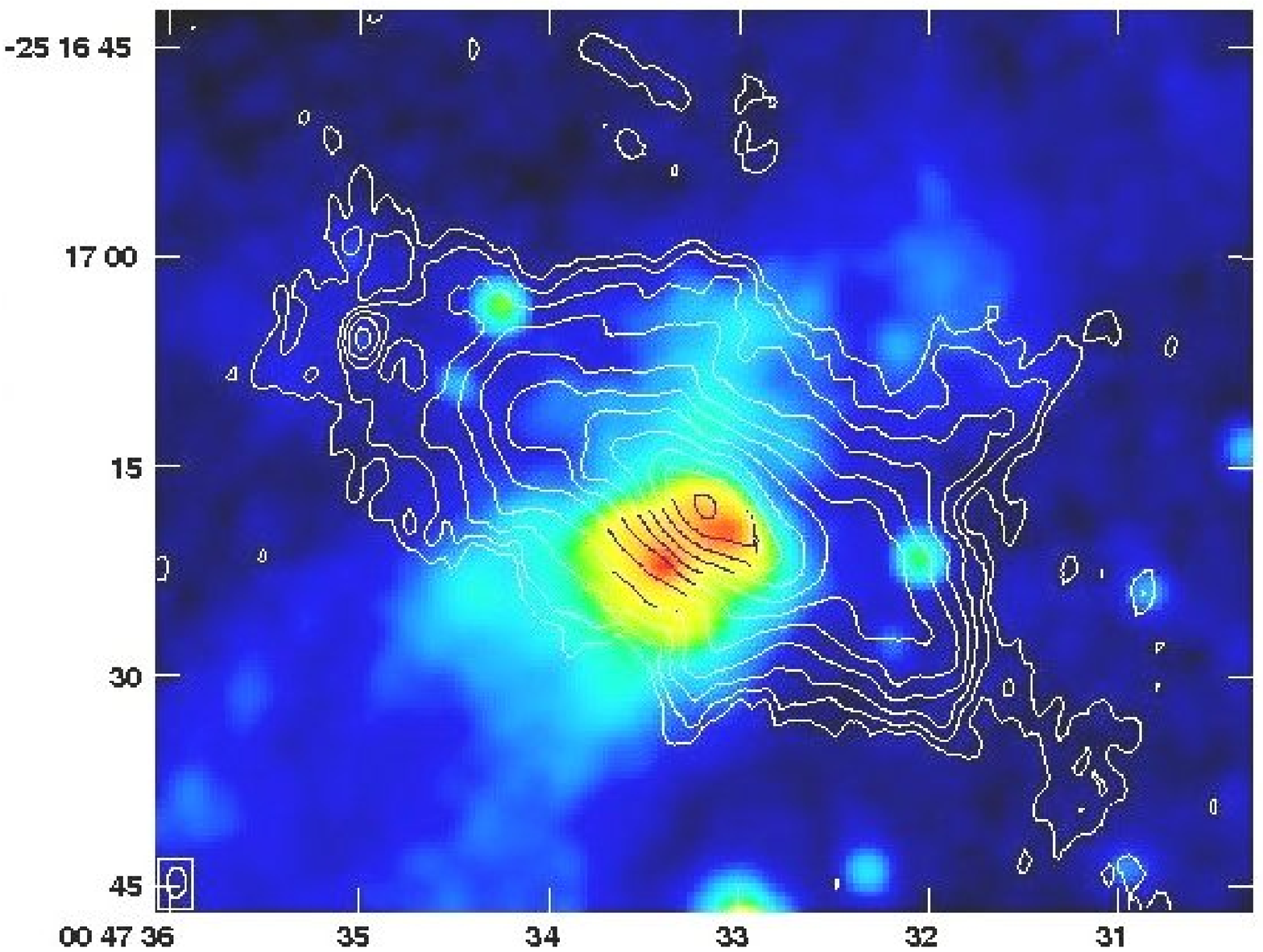}
\caption{Total radio emission (contours) of the inner starburst
region of the galaxy NGC~253, observed at 20~cm wavelength with the
{\it VLA}\ at a resolution of 1.3'' x 2.2'' (Heesen et al.
\cite{heesen11}). The colour image shows the X-ray emission
(0.5-5\,keV) measured with the {\it CHANDRA}\ satellite.}
\label{fig:n253_X}
\end{center}
\end{minipage}\hfill
\begin{minipage}[t]{6.5cm}
\begin{center}
\includegraphics[bb = 34 35 350 402,width=5.4cm,clip=]{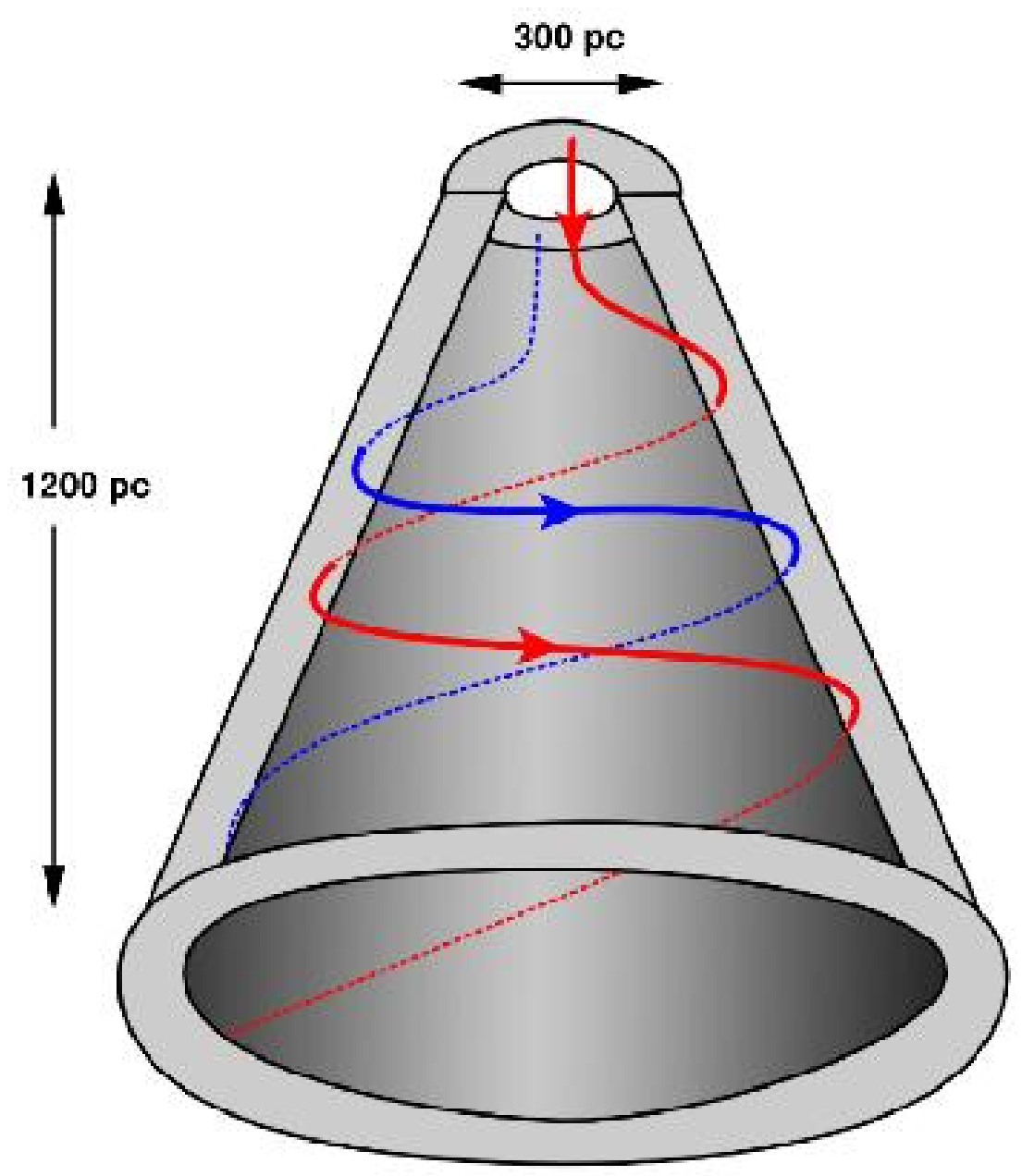} 
\caption{Model of the helical magnetic field in the walls of the
outflow cone (Heesen et al. \cite{heesen11}).}
\label{fig:n253_model}
\end{center}
\end{minipage}
\end{figure*}

Faraday rotation measures (RM) between 3\,cm and 6\,cm wavelengths
are consistent with an axisymmetric, even-parity, spiral magnetic
field in the disk at radii of more than 2\,kpc, while the field in
the inner disk is anisotropic (Heesen et al. \cite{heesen11}). Near
the SE minor axis, in the region of the X-ray outflow cone
(Fig.~\ref{fig:n253_X}), RM jumps between +300\,rad\,m$^{-2}$ and
-300\,rad\,m$^{-2}$. This is interpreted as the Faraday rotation of
the polarized emission from the background disk by the helical field
in the walls of the outflow extending to at least 1.2\,kpc from the
center (Fig.~\ref{fig:n253_model}). Only the inner part of this
helical field structure is seen in emission
(Fig.~\ref{fig:n253_center}). The northern outflow cone is located
behind the disk, so that its X-ray emission is mostly absorbed and
cannot Faraday-rotate the radio emission.

The galaxy NGC~3079 hosts an energetic wind-driven outflow which
extends 3\,kpc above and below the plane (Duric \& Seaquist
\cite{duric88}) and emits strong and highly polarized synchrotron
emission, in contrast to NGC~253. Radio-bright nuclear outflows seem
to be rare.

\section{Summary and outlook}

Magnetic fields in the central regions of galaxies are dynamically
important. Magnetic stress drives gas inflow from the circumnuclear
regions towards the nuclei. This process connects the star formation
activity to the accretion rate and could be important for the
formation of supermassive black holes. Ordered fields as observed by
polarized radio emission may trace the direction of gas in- and
outflows.

Results have been obtained so far only for radio-bright and nearby
galaxies. A better insight into the interaction between gas and
magnetic fields, especially near galactic nuclei, needs higher
resolution and higher sensitivity. The {\it EVLA}\ has largely
improved the sensitivity for radio continuum observations. The {\it
EVLA}\ project {\it CHANGES}\ (PI: Judith Irwin) will search for
outflows in a large sample of edge-on galaxies. The goal for the
next decade is the {\it Square Kilometre Array (SKA)}\ with a
collecting area of about 100x that of present-day telescopes. The
SKA will open a new era in the observation of cosmic magnetic fields
(Beck \cite{beck10}). Starburst galaxies will be observable out to
redshifts of $z\simeq3$ (Murphy \cite{murphy09}).

\end{document}